\documentclass[12pt,twocolumn]{iopart}

\usepackage{bm}
\usepackage{graphicx}
\usepackage{amssymb}

\newcommand{\itover}[2]{\,\hspace{-.15mm}#1{\!\hspace{.15mm}#2}}

\newcommand{\ithat}[1]{\itover{\hat}{#1}}

\newcommand{\zervec}{\bm{0}}

\newcommand{\deriv}[3]{\frac{#3\hspace*{-.06em} {#1}}{#3\hspace*{.06em} {#2}}}
\newcommand{\derivil}[3]{{#3 {#1}/#3 {#2}}}

\newcommand{\parder}[2]{\deriv{#1}{#2}{\partial}}
\newcommand{\parderil}[2]{\derivil{#1}{#2}{\partial}}

\newcommand{\EQ}{\begin{equation}}
\newcommand{\EN}{\end{equation}}
\newcommand{\EQA}{\begin{eqnarray}}
\newcommand{\ENA}{\end{eqnarray}}
\newcommand{\eq}[1]{(\ref{#1})}

\newcommand{\eqs}[2]{(\ref{#1}) and~(\ref{#2})}

\newcommand{\Sec}[1]{Sect.~\ref{#1}}

\newcommand{\Fig}[1]{Figure~\ref{#1}}
\newcommand{\FFig}[1]{Figure~\ref{#1}}
\newcommand{\Figs}[2]{Figures~\ref{#1} and \ref{#2}}

\newcommand{\Tab}[1]{Table~\ref{#1}}

\newcommand{\mean}[1]{\overline{ #1}}

{}
{}
\newcommand{\meanFFFF}{\overline{\mbox{\boldmath ${\cal F}$}}{}}{}
\newcommand{\meanFFFh}{\ithat{\meanFFF}}

{}
{}
\newcommand{\meanEMF}{\overline{\mbox{\boldmath ${\cal E}$}}{}}{}
{}
{}
{}
{}
{}
{}
{}
{}
{}
{}
{}
{}
{}
{}
{}
{}
{}
\newcommand{\meanUU}{\overline{\bm{U}}}

{}
{}
{}

\newcommand{\meanC}{\overline{C}}

\newcommand{\meanH}{\overline{H}}

\newcommand{\meanQ}{\overline{Q}}

\newcommand{\meanFFF}{\overline{\cal F}}

\newcommand{\meanRRR}{\overline{\cal R}}
\newcommand{\meanQQQ}{\overline{\cal Q}}

{}

{}
{}

\newcommand{\PC}{{\sc Pencil Code}~}
\newcommand{\xx}{\bm{x}}

\newcommand{\uu}{\mbox{\boldmath $u$} {}}
\newcommand{\UU}{\mbox{\boldmath $U$} {}}

\newcommand{\SSS}{\mbox{\boldmath $S$} {}}

\newcommand{\ff}{\mbox{\boldmath $f$} {}}

\newcommand{\FF}{\mbox{\boldmath $F$} {}}

\newcommand{\nab}{\mbox{\boldmath $\nabla$} {}}

\newcommand{\ii}{{\rm i}}

\def\Mu{\eta_{\scriptscriptstyle\cal E}}  
\def\taukapnu{\tau_{\nu\kappa}}
\def\taukapkap{\tau_{\kappa\kappa}}
\def\Cnukap{C_{\nu\kappa}}
\def\Ckapkap{C_{\kappa\kappa}}
\def \tauvisc{\tau_{\rm visc}}
\def\Sc{\mbox{\rm Sc}}

\def\Rey{\mbox{\rm Re}}
\def\Pe{\mbox{\rm Pe}}

\def\cs{c_{\rm s}}

\def\kf{k_{\rm f}}

\def\urms{u_{\rm rms}}

\def\etat{\eta_{\rm t}}

\begin{document}

\title[Mean-field closure parameters for passive scalar turbulence]{Mean-field closure parameters for passive scalar turbulence}

\author{J E Snellman$^{1,2}$, M Rheinhardt$^{1,2}$, P J K\"apyl\"a$^{1,2}$, M J Mantere$^{1,2}$, A Brandenburg$^{2,3}$}

\address{1. Department of Physics, Gustaf H{\"a}llstr{\"o}min katu 2a (PO Box 64), FI-00014 University of Helsinki, Finland}
\address{2. NORDITA, Roslagstullsbacken 23, SE-10691 Stockholm, Sweden}
\address{3. Department of Astronomy, Stockholm University, SE-10691 Stockholm, Sweden}
\ead{jan.snellman@helsinki.fi}

\begin{abstract}
  Direct numerical simulations of isotropically forced homogeneous
  stationary
  turbulence with an imposed passive scalar
  concentration
   gradient are compared with
  an analytical closure model 
  which provides
  evolution
  equations for the 
  mean
  passive scalar flux and variance.
  Triple correlations of fluctuations appearing in these
  equations are described in terms of relaxation terms proportional to
  the quadratic correlations. Three methods are used to extract the
  relaxation
   timescales $\tau_i$
   from direct numerical simulations.
   Firstly, we insert the closure
  ansatz into our equations, assume stationarity, and solve for
  $\tau_i$.
  Secondly, we use only the closure ansatz itself and
  obtain $\tau_i$ from the ratio of quadratic and triple
  correlations. Thirdly we remove the imposed passive scalar gradient
  and fit an exponential decay law to the solution. 

  We vary the Reynolds ($\Rey$) and P\'eclet ($\Pe$) numbers while keeping 
  their ratio 
  at unity
  and the
  degree of scale separation
  and
  find 
  for large $\Rey$
  fair correspondence between the
  different methods.
   The ratio of the turbulent relaxation 
  time of passive scalar flux 
  to the turnover time of turbulent eddies is of the order of
  three, which is in remarkable agreement with earlier work.
  Finally we make an effort to extract the relaxation timescales
  relevant for the viscous and diffusive effects. We find two regimes
  which are valid for small and large $\Rey$,
  respectively,
  but the dependence
  of the parameters
  on
  scale separation suggests that
  they
  are not universal.

\end{abstract}


\section{Introduction}
Fluid flows in astrophysical bodies are most often highly
turbulent. Direct modeling of such high-Reynolds-number flows is
currently impossible.
Consequently, greatly enhanced diffusivities or modified diffusion operators
are often applied in simulations \cite{BN11}. Such models are 
challenging in terms of
the required computational resources, so wide-ranging parameter studies
cannot be performed.

An alternative approach is to separate the large and small scales,
and derive equations for the former in which correlations of small-scale 
quantities are parameterized. 
This is usually referred to as mean-field theory, see e.g.\ 
\cite{M78,KR80,R89,RH04}.
Various schemes have been
introduced to close the equations
for the correlations of small-scale quantities.
In astrophysics, the second-order correlation approximation (SOCA), and the
minimal $\tau$ approximation (MTA, see e.g.\ \cite{BF02,BF03}) are
widely used.
The relevant relaxation time in MTA has been determined numerically
for passive scalar transport \cite{BKM04} as well as for the $\alpha$ effect
in mean-field electrodynamics \cite{BS05,BS07}.
Another approach, related to the MTA, where a relaxation term is
invoked to describe the higher order 
correlations
has been introduced in
\cite{O03}. In 
this `Ogilvie approach'
several nondimensional coefficients are
invoked to describe 
physically motivated parameterizations of
the higher order correlations in terms of relaxation and isotropization 
terms. This model has been applied to different
physical setups in order to calibrate the coefficients
\cite{GO05,MG07,LKKBL09,GOMS10}.

The validity of the various approximations can in principle be tested
by comparing with direct simulations in the same parameter regime. In
practise this is often not easy due to the limited parameter range
accessible by the simulations.
The starting point for such studies has been isotropically forced
homogeneous turbulence under the influence of rotation \cite{KB08} and/or
shear \cite{SKKL09}. In a recent work, the timescales related to
diffusion and isotropization that appear in the Ogilvie approach have
been studied \cite{SBKM11}. In the present study we extend the work of
\cite{SBKM11} to the passive scalar case.

\section{Mean-field modelling}\label{sec:mf}
\subsection{Ideal case}
\label{sec:ideal}
Let us consider the transport of a passive scalar under the influence of a turbulent fluid motion.
For simplicity we assume a homogeneous, incompressible fluid and neglect at first diffusion and viscous dissipation.
Then the governing equations for the concentration of the passive scalar, $C$, and the fluid velocity $\UU$ read
\begin{eqnarray}
\parder{C}{t} &=& -\nab \cdot (\UU C) = - \UU\cdot\nab C,\label{eq:Cbas}\\
\parder{\UU}{t} &=& - (\UU\cdot \nab) \UU -  \frac{1}{\rho} \nab P + \FF,\quad \nab\cdot\UU =0,\label{eq:Ubas}
\end{eqnarray}
where $P$ is the pressure, $\rho$ is the constant density and $\FF$ is
a forcing function with $\nab\cdot\FF=0$
(with the unit `force per mass').
Upon introduction of a Reynolds averaging procedure, indicated by an overbar, $C$ and $\UU$ are decomposed
into mean and fluctuating parts, $C=\meanC+c$, $\UU=\meanUU+ \uu$. The fluctuating fields, represented by lowercase letters, are then  
governed by
\begin{eqnarray}
\parder{ c}{t}  &=& - \uu \cdot \nab\meanC - \meanUU\cdot \nab c - (\uu \cdot \nab c)',\label{eq:Cfluct} \\
\parder{\uu}{t} &=& - (\uu \cdot \nab) \meanUU - (\meanUU \cdot \nab) \uu - \left((\uu \cdot \nab) {\uu}\right)' -  \frac{1}{\rho} \nab p + \ff,\label{eq:Ufluct}
\end{eqnarray}
where the prime indicates extraction of the fluctuating part, e.g.,  $(\uu c)' = \uu c - \overline{\uu c}$.
Simplifying further, we stipulate the absence of a mean velocity $\meanUU$ and  assume that the forcing has no mean part, i.e.,
$\FF=\ff$.
In the present case, the goal of mean-field modeling consists in deriving a closed equation for the mean concentration $\meanC$. From \eq{eq:Cbas} 
and \eq{eq:Cfluct}, together with $\meanUU=\zervec$ we obtain directly
\begin{equation}
\parder{\meanC}{t} = -\nab \cdot \meanFFFF \label{eq:Cmean} 
\end{equation}
with the mean density of the passive scalar flux, $\meanFFFF=\overline{c\uu}$.
So the task of closing \eq{eq:Cmean} reduces to representing  $\meanFFFF$ by the mean concentration $\meanC$.
In the standard mean-field approach,  \eq{eq:Cfluct} is solved for a prescribed fluctuating velocity $\uu$, usually under some
simplifying assumptions which inevitably limit the applicability of the obtained results. The solution is employed
to express $\meanFFFF$ in terms of $\meanC$.
Alternatively, one can abstain from deriving such an explicit solution for the fluctuating concentration $c$ and instead strive for
establishing an {\em evolution equation} for $\meanFFFF$ which of course again has to be closed in the sense that the
only variables occurring are the mean quantities $\meanC$ and
$\meanFFFF$ themselves.
Such an equation is obtained by multiplying \eq{eq:Cfluct} with $\uu$ and \eq{eq:Ufluct} with $c$, summing up and averaging, arriving at
\begin{equation}
\parder{\meanFFFF}{t} = -\overline{\uu\nab \cdot(\uu \meanC)} -\overline{\uu\nab \cdot(\uu c)} - \overline{c\left((\uu\cdot\nab)\uu\right)} 
                                             -\frac{1}{\rho}\,\overline{c\nab p} + \overline{c\ff}. \label{eq:Fmean}
\end{equation}
By virtue of the incompressibility of the fluid the fluctuating pressure $p$ can be expressed by the velocity fluctuations:
\EQ
    \nab^2 p = - \rho\left(\parder{u_i}{x_j}\parder{u_j}{x_i}\right)'   \label{eq:pPoiss}
\EN
which for an infinitely extended medium and vanishing $p$ at infinity is readily solved by 
\EQ
      p = \frac{\rho}{4\pi}\int \frac{\left(\parderil{u_i}{x_j}\parderil{u_j}{x_i}\right)'(\xx')}{|\xx-\xx'|}d\,^3 x' .   \label{eq:pSolu}
\EN
Now we can conclude that the second, third, and fourth terms on the
r.h.s. of \eq{eq:Fmean} are quadratic in $\uu$ and linear in $c$,
hence represent third order correlations.
Following \cite{GOMS10} we introduce here the {\em closure assumption} 
\EQ
-\overline{\uu\nab \cdot(\uu c)} - \overline{c\left((\uu\cdot\nab)\uu\right)} 
                                             -\frac{1}{\rho}\,\overline{c\nab p}  = - \frac{\meanFFFF}{\tau_6}  \label{eq:tau6}
\EN
with a {\em relaxation time} $\tau_6$.
Upon further neglect of the correlation  $\overline{c\ff}$, we arrive at
\EQ
\parder{\meanFFF_i}{t} = -\overline{u_i u_j}\,\parder{\meanC}{x_j} - \frac{\meanFFF_i}{\tau_6} = - \meanRRR_{ij} \parder{\meanC}{x_j} - \frac{\meanFFF_i}{\tau_6}, \label{eq:FmeanTau}
\EN
which governs the evolution of $\meanFFFF$.
Here, $\meanRRR_{ij}=\overline{u_i u_j}$ stands for the Reynolds stress tensor.
Since a {\em passive} scalar would not
act back onto the velocity, $\meanRRR_{ij}$ can here
be considered given and the closure is completed.
Nevertheless, in the presence of rotation, shear, or gravity,
\eq{eq:Ufluct} contributes quadratic correlations to \eq{eq:Fmean}
even in the kinematic case; see, e.g., \cite{KB08,SKKL09,MB10}.
Equation \eq{eq:FmeanTau} is similar to the penultimate row of
Eq.~(53) in \cite{GOMS10} when replacing temperature perturbation
$\Theta$ by $C$ and neglecting the buoyancy term.
Note that for small $\tau_6$, that is, fast  relaxation, $\meanFFF_i$ will follow the inhomogeneity in \eq{eq:FmeanTau} almost instantaneously, hence
\EQ
      \meanFFF_i \approx -\tau_6  \meanRRR_{ij}\,\parder{\meanC}{x_j},  \label{eq:FmeanInst}
\EN
and we may interpret $\tau_6  \meanRRR_{ij}$ as a {\em turbulent diffusivity tensor}. For a discussion of its relationship to traditional mean-field results, see \Sec{sect:comp}.

To facilitate further comparisons to \cite{GOMS10,BHB11}, where an
additional evolution equation for the mean temperature perturbation variance
$\overline{\Theta^2}$ is derived, we give here an analogous equation
for $\overline{c^2}= \meanQQQ $, although it is not necessary for
completing the closure:
\EQ
\parder{\meanQQQ}{t} = -2 \overline{c\uu}\cdot\nab \meanC - 2\overline{c(\uu \cdot \nab c)}. \label{eq:Qmean}
\EN
We note in passing that the $\meanQQQ$ term becomes important
in reacting flows \cite{BHB11}.
Setting 
\EQ 
- 2\overline{c(\uu \cdot \nab c)}= -\meanQQQ/\tau_7  \label{eq:tau7}
\EN
 with another relaxation time $\tau_7$, the closed
equation reads
\EQ
\parder{\meanQQQ}{t} = -2 \meanFFFF\cdot\nab \meanC - \frac{\meanQQQ}{\tau_7} \label{eq:QmeanTau}
\EN
and we have full analogy to the last row of Eq.~(53) in \cite{GOMS10}.

Until now we have not constrained the properties of the turbulence, in particular we have not required isotropy or homogeneity.
For example, inhomogeneous turbulence could be thought of
giving rise to position-dependent relaxation times.
However, from a strict point of view, the $\tau$ ansatz \eq{eq:tau6} is 
only consistent with an isotropic or uni-axial $\uu$ turbulence where the preferred direction of the latter coincides with the direction of $\meanFFFF$.
Consequently, turbulent properties of $\uu$ must not change along any other direction.  
The same restrictions should of course hold for the concentration fluctuations $c$, but this is in conflict with the presence of the second preferred
direction $\nab \meanC$ in this turbulence. Hence, \eq{eq:tau6} can be
strictly justified only under very specific circumstances.

For that reason and for the sake of simplicity,
we specify now the mean as horizontal average, i.e., as average over
all $x$ and $y$.
Consequently, all mean quantities depend on $z$ only
and only the $z$ component of $\meanFFFF$ is relevant.
If we further restrict $\uu$ to have at best a $z$ anisotropy, then
there is only a single preferred direction, namely that of $\meanFFFF$
and the ansatz \eq{eq:tau6} is legitimate.
The system of mean field equations then simplifies
to 
\EQ
\parder{\meanC}{t} = -\parder{\meanFFF_z}{z}, \quad
\parder{\meanFFF_z}{t} = - \meanRRR_{zz} \parder{\meanC}{z} - \frac{\meanFFF_z}{\tau_6},\quad
\parder{\meanQQQ}{t}  = -2 \meanFFF_z\parder{\meanC}{z}- \frac{\meanQQQ}{\tau_7} \,.  \label{eq:mean1D}
\EN
where $ \meanRRR_{zz}$, $\tau_6$ and $\tau_7$ could still depend on $z$ and $t$.
Assuming now further homogeneous and statistically stationary fluctuations
$\uu$ and $c$ and a uniform gradient of $\meanC$, $\nab \meanC=(0,0,G)$,
a stationary regime of \eq{eq:mean1D} should be given by
\EQ
\meanFFF_z = \mbox{const.} = -\tau_6 \meanRRR_{zz} G,\quad  \meanQQQ = -2 \tau_7 \meanFFF_z G =   2 \tau_6 \tau_7 \meanRRR_{zz} G^2. \label{eq:meanStat}
\EN
Let us now assume that in a direct numerical simulation (DNS)
Eqs.~\eq{eq:Cbas} and \eq{eq:Ubas} with an appropriately defined forcing
$\ff$ and an imposed uniform $G$
are integrated in time until a statistically stationary
regime is established. Extracting now all mean quantities occurring in \eq{eq:meanStat} from the numerical solution and assuming validity of the model \eqs{eq:FmeanTau}{eq:QmeanTau},
it is obviously possible to determine the crucial relaxation times $\tau_{6,7}$ from such runs (method M1). On the other hand, $\tau_{6,7}$ should of course also obey their
defining relations \eq{eq:tau6} and \eq{eq:tau7}. The third-order correlations $\overline{\uu\nab\cdot(\uu c)}$, $\overline{c\left((\uu\cdot\nab)\uu\right)} $, 
$\overline{c\nab p}$ and $\overline{c(\uu \cdot \nab c)}$ are again accessible in the DNS results and open up an independent 
path for determining the relaxation times (method M2). At the same time, it can also be checked to what extent the neglect of $\overline{c\ff}$ is justified.

\begin{figure}[t!]\begin{center}
\includegraphics[width=\linewidth]{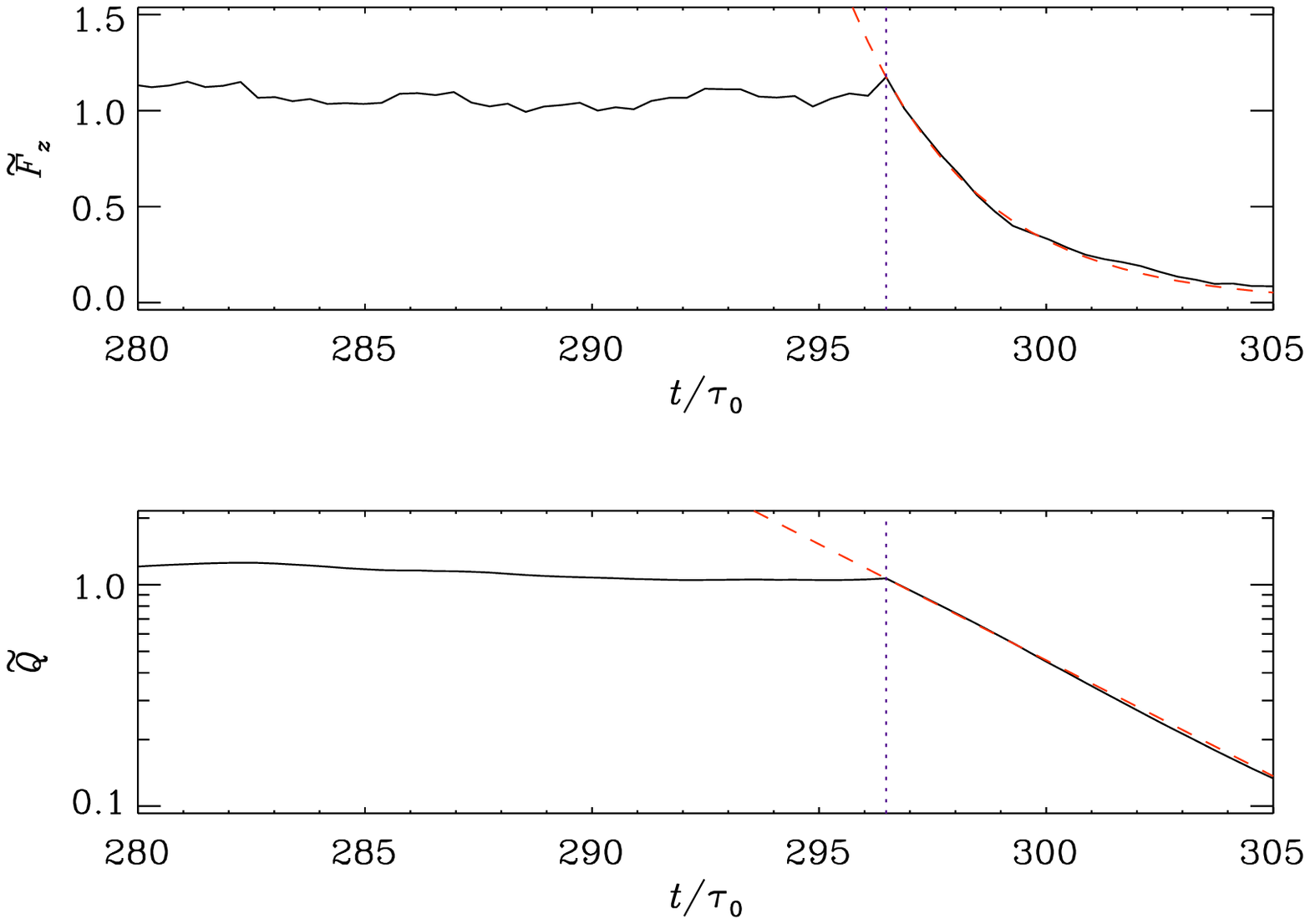}
\end{center}\caption[]{ \label{fig:dec}
Method M3 of estimating the timescales $\tau_{6,7}$
from decaying $\meanFFF_z$ and $\meanQQQ$ 
normalized by their mean values in the stationary state:
$\widetilde{\meanQQQ}=\langle\meanQQQ\rangle_{z}/\langle\meanQQQ\rangle_{zt}$, $\widetilde{\meanFFF}_z=\langle\meanFFF_z\rangle_{z}/\langle\meanFFF\rangle_{zt}$, where $\langle\rangle_\xi$ 
denotes averaging over the variable $\xi$.
Dotted vertical line: switching off of the imposed mean concentration gradient $G$.
$\tau_0=(\urms\kf)^{-1}$ -- dynamical time.
Dashed red lines: fit by exponentials in $t/\tau_0$.
}\end{figure}

Another approach to extract $\tau_{6,7}$ is available from decay
experiments, for which, after having reached a stationary state in the
DNS, the imposed gradient of $\meanC$
is switched off. Then, according to \eq{eq:FmeanTau} and \eq{eq:QmeanTau}, $\meanFFFF$ and $\meanQQQ $ should decay uniformly in space and exponentially in time with the increment $\tau_6^{-1}$ and $\tau_7^{-1}$, respectively, and  can be identified with the 
decay rates measured in the DNS (method M3).
An exemplification of this method is shown in \Fig{fig:dec}.

The goal of this paper consists in systematically testing the validity
of the presented closure assumptions for a range of Reynolds and
P\'eclet numbers as well as different
levels of scale separation. From this we expect hints with respect to
the validity of the Garaud model of turbulent convection
\cite{GOMS10}.
 
\subsection{Non-ideal effects}
\label{sec:nonideal}
Admitting now diffusion and viscous dissipation, we have to add the term $\kappa \nab^2 C$ with the diffusivity $\kappa$ on the r.h.s. of \eq{eq:Cbas} and $\nu \nab^2\UU$ 
with the kinematic viscosity $\nu$ on the r.h.s. of \eq{eq:Ubas}. Consequently, in the evolution equation \eq{eq:Fmean} for $\meanFFFF$,  the additional terms $\nu\overline{c\nab^2\uu}$ and
$\kappa\overline{\uu\nab^2c}$ occur on the r.h.s.. Rewriting their sum as
\EQ
\nu\nab^2\meanFFF_i - 2 \nu \overline{\parder{c}{x_j} \parder{u_i}{x_j}} + (\kappa-\nu)\overline{u_i \nab^2 c}   \label{eq:diff1}
\EN
or more symmetric, as done in \cite{GOMS10}, as
\EQ
\frac{\nu+\kappa}{2}\nab^2\meanFFF_i - (\nu+\kappa) \overline{\parder{c}{x_j} \parder{u_i}{x_j}} + \frac{\nu-\kappa}{2}\overline{\parder{}{x_j}\left(c\parder{u_i}{x_j} - u_i \parder{c}{x_j}\right)}    \label{eq:diff2}
\EN
does nevertheless not allow a representation entirely by the mean flux.
Even in the (very particular) case $\kappa=\nu$ the second terms of
\eqs{eq:diff1}{eq:diff2} remain.
As a skyhook, the second and third terms are replaced by the
$\tau$-ansatz-like expression, $-\meanFFFF/\taukapnu$ although they
contain second order rather than third order correlations.
Analogously, on the r.h.s. of \eq{eq:Qmean}, diffusion requires a term 
\EQ
2\kappa\overline{c\nab^2 c} = \kappa \nab^2\meanQ - 2 \kappa\overline{(\nab c)^2}  \label{eq:diffQ}
\EN
and the second term is replaced by $-\meanQ/\taukapkap$.
Note that diffusion of $\meanFFFF$ and $\meanQQQ$ modelled in this way is obviously  determined by the {\em molecular} (or microscopic) diffusivities. 

In astrophysical applications the deviation from ideal conditions is usually small, and quantities expressing this smallness are given by the
Reynolds and P\'eclet numbers, $\Rey$ and $\Pe$, which reflect the strength of advection relative to diffusion:
\EQ
  \Rey = \urms \ell/\nu, \quad \Pe = \urms \ell/\kappa,  \label{eq:numdef}
\EN
where $\ell$ is a characteristic scale of the turbulence. We will further make use of the Schmidt number $\Sc=\nu/\kappa=\Pe/\Rey$.

\subsection{Summary of method M2}
So, summarizing all the terms used to determine $\tau_{6\nu\kappa}$ and 
$\tau_{7\kappa\kappa}$ from Method M2 (ideal and non-ideal, see equations 
\eq{eq:tau6}, \eq{eq:tau7}, \eq{eq:diff2} and \eq{eq:diffQ}), we have
\begin{eqnarray}
 \frac{\meanFFF_z}{\tau_{6\nu\kappa}} = \overline{u_z\nab \cdot(\uu c)} + \overline{c\left((\uu\cdot\nab)u_z\right)} + \frac{1}{\rho}\,\overline{c\,\nab_{\!z} \,p} + \nonumber\\
+ (\nu+\kappa) \overline{\nab c \cdot\nab u_z} - \, \frac{\nu-\kappa}{2}\,\overline{\left(c\nab^2{u_z} - u_z \nab^2 c\right)} ,
\label{eq:m2ni1}
\end{eqnarray}
\EQ
\frac{\meanQQQ}{\tau_{7\kappa\kappa}} = 2\left(\overline{c(\uu \cdot \nab c)} + \kappa\overline{(\nab c)^2}\right).
\label{eq:m2ni2}
\EN

\subsection{Scaling of the relaxation times}
\label{sec:scal}
For the relaxation times $\tau_{6,7,\nu\kappa,\kappa\kappa}$ some reasonable scaling assumptions are in order and we follow essentially the choices of \cite{GOMS10}:
$\tau_{6,7}$, belonging to third-order correlation terms, are expressed as  $(C_{6,7} \urms k_1)^{-1}$, and $\tau_{\nu\kappa,\kappa\kappa}$, belonging to diffusive
second-order correlation terms are written as 
\EQ
(\Cnukap(\nu+\kappa)k_1^2/2)^{-1} \quad \mbox{and} \quad (\Ckapkap\kappa k_1^2)^{-1},  \label{eq:scal}
\EN
respectively. 
The first of these expressions seems to be appropriate only for $|\nu-\kappa|\ll\nu+\kappa$, hence in general the scaling ansatz should read instead
\EQ
\Big(\big(C_{\nu+\kappa}\frac{\nu+\kappa}{2}+C_{\nu-\kappa}(\nu-\kappa)\big)k_1^2\Big)^{-1}\!\!\!. \label{eq:fit_Scne1}
\EN
Here $k_1=2\pi/L$ is the smallest wavenumber consistent with the box size, $L$.
The crucial question is now:
are the constants $C_{6,7,\nu\kappa,\kappa\kappa}$ universal, at least
for a given type of turbulence, and in particular, are they independent of
the dimensionless numbers  of the problem, i.e., $\Rey$ and $\Pe$, and of the degree of scale separation?
A preliminary answer to this question was given in \cite{BKM04}, where
the timescales were found to show a slightly increasing trend with
increasing scale separation (see their Fig.~4).

Methods M1 and M3 for determining the relaxation times described in \Sec{sec:ideal} have now to be modified in the following way:
In \eq{eq:meanStat} we have to replace $\tau_6$ by $\tau_6\taukapnu/(\tau_6+\taukapnu)\equiv\tau_{6\nu\kappa}$ and $\tau_7$ by  $\tau_7\taukapkap/(\tau_7+\taukapkap)\equiv\tau_{7\kappa\kappa}$. Both methods then
deliver only these aggregates and we have to employ the different scalings of the relaxation times to figure out the individual constants $C_\ast$.
In contrast, method M2 has merely to be extended to include also the additional  second-order correlations showing up in \eq{eq:diff2} and \eq{eq:diffQ},
that is, to use expressions (\ref{eq:m2ni1}) and (\ref{eq:m2ni2}).

\subsection{Comparison with traditional results}
\label{sect:comp}
A standard mean-field approach to \eq{eq:Cbas}, employing SOCA, that is,
neglecting  $(\uu\cdot\nab c)'$ in \eq{eq:Cfluct} yields straightforwardly
\EQ
      \meanFFF_i(t) = -\int_0^{\infty} \overline{u_i(t) u_j(t-\tau)} \,\parder{\meanC}{x_j}(t-\tau)\, d\tau \label{eq:SOCA}
\EN
from which, under the assumption of good {\em temporal scale separation},
\EQ
      \meanFFF_i(t)= -\int_0^{\infty} \overline{u_i(t) u_j(t-\tau)} \,d\tau \,\parder{\meanC}{x_j}(t) = -\kappa_{ij} \parder{\meanC}{x_j}  \label{eq:SOCAtsep}
\EN
can be concluded. $\kappa_{ij} = \int_0^{\infty} \overline{u_i(t)
u_j(t-\tau)}\, d\tau = \tau_{\rm c} \overline{u_i(t) u_j(t)}$ can readily
be identified as turbulent diffusivity tensor.
Here the correlation time $\tau_c$ is just defined by the last
identity.
This clearly resembles the result \eq{eq:FmeanTau} with $\tau_6$
being identified with
the correlation time $\tau_c$, the more so as for the validity of both
\eq{eq:SOCAtsep} and \eq{eq:FmeanTau} the relevant time parameter has,
in a sense, to be small.

Relaxing the assumption of good temporal scale separation, that is
retaining \eq{eq:SOCA}, we observe the presence of the so-called
{\em memory effect} \cite{HB09},
that is, the influence of $\parderil{\meanC}{x_j}$ at earlier times $t-\tau$ on the mean flux at time $t$ by virtue of a convolution. Performing a Fourier transform with respect  to time, this convolution turns into a simple multiplication and we can write
\EQ
    \meanFFFh_i(\omega) = -\ithat{\kappa}_{ij}(\omega) \ithat{G}(\omega)
\EN
with a frequency-dependent turbulent diffusivity tensor $\ithat{\kappa}_{ij}$. This quantity is directly accessible to the test-field method as described, e.g., in \cite{HB09}.
Based on numerical simulations, and without resorting to SOCA, it has
been found that for homogeneous isotropic turbulence a satisfactory
approximation is accomplished already by
\EQ
   \hat{\kappa}_{ij}(\omega) = \delta_{ij}\frac{\kappa_0}{1-\ii \omega\tau_\kappa} \label{eq:fit1}
\EN
where $\kappa_0$ is the turbulent diffusivity for stationary fields and $\tau_\kappa$ is independent of $\omega$.
A slightly better fit is provided by 
\EQ
   \hat{\kappa}_{ij}(\omega) = \delta_{ij}(1+K)\kappa_0\frac{1-\ii \omega\tau_\kappa}{(1-\ii \omega\tau_\kappa)^2+K} \label{eq:fit2}
\EN
with a constant $K$.
Turning back to the physical space the first approximation \eq{eq:fit1} is equivalent to 
\EQ
    \meanFFFF + \tau_\kappa \parder{\meanFFFF}{t} = - \kappa_0 \nab \meanC
\EN
or
\EQ
    \parder{\meanFFFF}{t} =  - \frac{\kappa_0}{\tau_\kappa} \nab \meanC - \frac{\meanFFFF}{\tau_\kappa}.  \label{eq:TFtsep}
\EN
Again, there is striking similarity to \eq{eq:FmeanTau}. Thus by comparing $\tau_6$ to numerical results for $\tau_\kappa$, a further independent way of checking \eq{eq:tau6} is provided.
The second fit \eq{eq:fit2} gives \cite{HB09}
\EQ
    (1+K)\meanFFFF + 2\tau_\kappa \parder{\meanFFFF}{t} + \tau_\kappa^2  \parder{\,^2 \meanFFFF}{t^2} = - (1+K)\kappa_0 \nab \left(\meanC + \tau_\kappa  \parder{\meanC}{t}\right)\label{eq:TFtsep1}
\EN
indicating the potential importance of higher temporal and mixed temporal/spatial derivatives.
Note that \eqs{eq:TFtsep}{eq:TFtsep1} are only valid for perfect scale separation in space.

For the general case of imperfect scale separation both 
with respect to space and time, we refer here to \cite{RB11}, albeit this work deals with the mean electromotive force of MHD rather
than with the mean flux of passive scalar transport.
In that work, non-locality due to imperfect spatial scale separation
shows up in the form
of a diffusion term $\Mu \nab^2 \meanEMF$ in the evolution equation for
$\meanEMF$ with a diffusivity $\Mu$  occasionally even larger than,
but of the order of
the SOCA estimate of the turbulent diffusivity in the high-conductivity
limit, $\etat = \tau_{\rm c}\urms^2/3$.
Clearly, this value can be very different from the molecular diffusivity.
When comparing with the diffusion term for $\meanFFFF$ identified in
\eq{eq:diff1} or \eq{eq:diff2} where only the microscopic diffusivities
occur we have to state that the Ogilvie approach
deviates in this respect significantly from what we expect from the traditional approach.
To reconcile them,  possible diffusion terms $\sim\nab^2\meanFFFF$ and $\sim\nab^2\meanQQQ$ had to be taken into account in the
parameterizing  ansatzes.

\subsection{Significance of method comparisons}
\label{sect:assess}
Let us finally discuss what is really ``tested" by comparisons of the
 results of the
 different methods M1--M3.
For an incompressible fluid with the specific conditions of our model
and under the assumption that no higher than the first order temporal derivative occurs,
an ansatz for $\meanFFFF$  analogous to \eq{eq:meanStat},
\EQ
   \parder{\meanFFF_z}{t} = - K_1 G - K_2 \meanFFF_z,
\EN
is exhaustive, as 
\begin{enumerate}
\item 
$c$ and hence $\meanFFFF$  is quite generally a linear and homogeneous functional of $\nab \meanC$ 
(even when the correlation $\overline{c\ff}$ is not neglected), and 
\item $G$ and hence 
$\meanFFF_z$ are spatially constant, both in the statistically steady state and during the decay of $\meanFFF_z$.
 (That is why the diffusive term $\propto\nab^2 \meanFFF_z$ is absent.)
\end{enumerate}
Note, however, that $K_1= \meanRRR_{zz}$ as in \eq{eq:meanStat}
{\em is} an assumption because a contribution proportional to $\nab C$
or $\parderil{\meanFFF_z}{t}$ can also be provided by the triple
correlations, the diffusive terms or by $\overline{c\ff}$.

Since the {\em passive} scalar $C$ does not influence the turbulent
velocity, the coefficients $K_{1,2}$ are completely determined by $\uu$
and hence true constants.
Consequently, any comparison of the methods M1, M2, and M3 tests the
influence of
 \begin{enumerate}
\item the weak compressibility of the fluid in our simulations, and
\item deviations of the simulated velocity turbulence from homogeneity, isotropy and statistical stationarity.
\end{enumerate}
Note, that the first influence can in principle be made arbitrarily small by increasing the sound speed $\cs$ in the numerical model,
likewise the second by increasing scale separation and extending time
ranges for averaging.

A comparison of M1 and M3 tests in addition the justification of
\begin{enumerate}
\item  the neglect of higher temporal derivatives of $\meanFFF_z$,
\item  the assumption $K_1= \meanRRR_{zz}$ which was employed in calculating $\tau_{6\nu\kappa} = 1/K_2$ from the steady state.
\end{enumerate}

On the other hand, a comparison of M1 and M2 tests again the assumption
$K_1= \meanRRR_{zz}$ and specifically to what extent the neglect of the
correlation $\mean{c\ff}$ is legitimate.
This affects $\tau_{6\nu\kappa}$ only, so we expect that here
the discrepancies between M1 and M2 are more pronounced than in
$\tau_{7\kappa\kappa}$.

With respect to the evolution equation for $\meanQQQ$ in \eq{eq:mean1D},  the same conclusions hold true as far as only the steady state
and the free decay with $G=0$ are taken into account.
However, considering a general transient as a consequence of switching $G$ between two non-zero constants, a richer
behavior should appear which perhaps allows further reaching conclusions.

\section{Numerical setup}
In order to take benefit of the capabilities of the \PC\footnote{freely available at  http://pencil-code.googlecode.com/} we solve
instead of the incompressible system \eqs{eq:Cbas}{eq:Ubas} the corresponding  equations for a compressible, but isothermal fluid\\[-.5mm]
\begin{eqnarray}
\parder{\,\UU}{t} &=& - \UU \cdot \nab\UU - \nab H + {\ff} + 2\nu\left(\nab\cdot\SSS(\UU)+ \SSS(\UU)\cdot\nab H/\cs^2 \right)  \label{eq:Ccomp}\\
\parder{H}{t} &=& - \UU\cdot\nab H -\cs^2\nab \cdot \UU,  \label{eq:Ucomp}\\
\parder{\,C}{t} &=& -\nab\cdot(C\UU) + \kappa\nab^2 C,   \label{eq:Ccomp}
\end{eqnarray}\\
where we employ the pseudo enthalpy $H=\cs^2\ln \rho$ instead of the
density; $\cs$ is the constant speed of sound and $ \SSS(\UU)$ the trace-less rate-of-strain
tensor $S_{ij} = (\parderil{U_i}{x_j}+\parderil{U_j}{x_i})/2-\delta_{ij}\nab\cdot\UU/3$.
Interpretation of the results of such simulations in terms of the incompressible model of course requires to keep the Mach number
$\urms/\cs$ small, typically $<0.1$.
Then it is particularly justified to replace the correlation
$\mean{c\nab p}/\rho$, included in the $\tau$ ansatz, by $\mean{c\nab h}$.

The equations are solved by equidistant sixth order finite differences in
space and an explicit third order time stepping scheme
with step size control for stability.
The computational domain is a cube with dimension $(2\pi)^3$ and grid resolutions ranging from
$32^3$ to $256^3$ according to the requirements raised by the values of
the Reynolds and P\'eclet numbers and the forcing wavenumber.
Boundary conditions are periodic throughout.
The fluctuating force $\ff$ is specified such that it generates an approximately homogeneous, isotropic and statistically stationary 
fluctuating velocity $\uu$. During any integration timestep, $\ff$ is a frozen-in linearly polarized (i.e., non-helical) plane wave with a
wave-vector which is consistent with the periodic boundary conditions and whose modulus is close to a chosen average value $\kf$.
The wave amplitude is kept fixed whereas the wavevector is randomly
changing between time steps and hence $\ff$ is approximately
$\delta$-correlated in time.  For further details, see \cite{B01}.

\section{Results}

\begin{table}
\caption{\label{tab:xnbres}Results from methods M1, M2, and M3 for different values of $s=\kf/k_1$. $\nu=\kappa$, $\Sc=1$.}
\begin{tabular*}{\textwidth}{@{}cc@{\hspace{5mm}}c@{\hspace{3mm}}c@{\hspace{5mm}}c@{\hspace{3mm}}c@{\hspace{4mm}}c@{\hspace{3mm}}c@{}}          
\br
 $s$ & $\rm Re$ & \multicolumn{2}{c}{M1} &   \multicolumn{2}{c}{M2} &  \multicolumn{2}{c}{M3} \\
         &                  & $\tau_{6 \nu \kappa}$ & $\tau_{7 \kappa \kappa}$ & $\tau_{6 \nu \kappa}$ & $\tau_{7 \kappa \kappa}$ & $\tau_{6 \nu \kappa}$ & $\tau_{7 \kappa \kappa}$ \\
\mr
1.5 & 51.02-552.47 & 2.09-2.42 & 4.14-4.28 & 2.36-2.71 & 4.10-6.86 & 1.85-2.96 & 3.64-5.21 \\
3 & 23.44-694.63 & 2.12-2.64 & 4.29-5.03 & 2.15-2.70 & 4.29-14.21 & 1.94-2.81 & 3.94-4.96 \\
 5 & 12.62-415.00 & 2.29-2.77 & 4.54-5.89 & 2.31-2.79 & 4.54-41.27 & 2.43-2.77 & 4.93-5.67 \\
 8 & 6.87-256.50 & 2.40-2.86 & 5.59-9.13 & 2.41-2.88 & -21.54-34.81 & 2.25-2.85 & 5.26-7.70 \\
10 & 5.06-204.34 & 2.09-2.87 & 6.00-8.06 & 2.10-2.93 & -10.43-8.93 & 2.24-2.99 & 3.44-4.50 \\
\br
\end{tabular*}
\end{table}
In general, the parameter space is spanned by the dimensionless numbers
$s=\kf/k_1$ (degree of scale separation), $\Rey$ and $\Pe$, in whose
definitions \eq{eq:numdef}
we specify the characteristic length $\ell$ for simplicity by $1/\kf$. 
In the following, we will however restrict ourselves to $\Rey=\Pe$, that is, $\Sc=1$ and leave the more general cases to future work.
By this, we avoid in particular the complication with the scaling ansatz \eq{eq:scal} which occurs for $\nu\ne\kappa$.
For methods M1 and M2, all averaged quantities were derived from the
simulations by performing in addition to the $xy$ averaging and a  temporal
averaging over an interval in which in particular the correlation
$\overline{c^2}$ was found to be statistically stationary.
Statistical errors were estimated by dividing this interval into 
three equally long parts and calculating averages over each of them.
These individual averages were compared to the average over the whole
interval, and the largest deviation was taken for the error estimate.

\begin{figure}[t!]\begin{center}
\includegraphics[width=\linewidth]{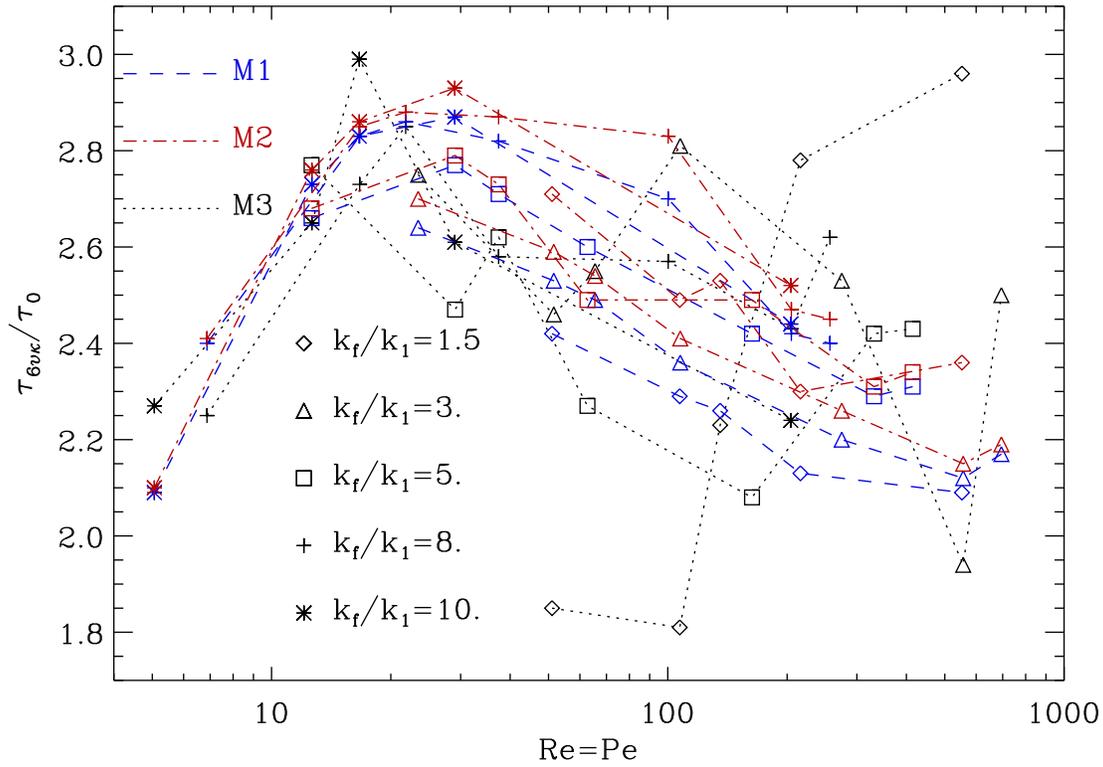}
\end{center}\caption[]{ \label{fig:rect6}
Comparison of relaxation timescales $\tau_{6 \nu \kappa}$, normalized by
the dynamical time $\tau_0=(\urms k_f)^{-1}$, as functions of the 
Reynolds/P\'eclet number from methods M1--M3.
Different symbols refer to different scale separations $s=\kf/k_1$ as indicated in the legend.
}\end{figure}

We performed a number of simulations, and extracted $\tau_{6 \nu \kappa}$ and 
$\tau_{7 \kappa \kappa}$ using methods M1, M2 and M3.
The results are summarized in \Tab{tab:xnbres}, where the several simulations 
are grouped together into  different sets 
by the values used for the scale separation $s$.
Within each set the 
Reynolds numbers were varied by changing $\nu$ ($=\kappa$). 
The timescales 
$\tau_{6 \nu \kappa}$ and $\tau_{7 \kappa \kappa}$ listed 
in \Tab{tab:xnbres} are also illustrated in  
\Figs{fig:rect6}{fig:rect7}, respectively.
We see that all three methods give quite similar results, with  
\begin{eqnarray}
\tau_{6 \nu \kappa}/\tau_0 \approx 2\ldots 3,\hspace{0.5cm} \mbox{and} \hspace{0.5cm} \tau_{7 \kappa \kappa}/\tau_0\approx 7 \ldots 10.
\end{eqnarray}
There are some exceptions,
however. When $s=1.5$, methods M2 and M3 yield 
$2\lesssim \tau_{7 \kappa \kappa}/\tau_0 \lesssim 3$, 
unlike method M1 which stays in range given above.
Moreover, although method M2 always yields positive values
for $\tau_{6 \nu \kappa}$, those for $\tau_{7 \kappa \kappa}$
are sometimes negative.  
The absolute values
 of the M2 results
can also be very 
high. This is because the sum of the correlations used 
to calculate $\tau_{7 \kappa \kappa}$ may become very small. 
This problem manifests itself mainly in the high Reynolds number runs.

\subsection{Universality of the closure ansatz}
As explained in \Sec{sec:nonideal}, methods M1 and M3 yield only the quantities $\tau_{6\nu\kappa}$ and
$\tau_{7\kappa\kappa}$. 
from which
the constants $C_{6,7,\nu\kappa,\kappa\kappa}$ can be extracted as follows: Recalling
the scalings \eq{eq:scal} we have:
\EQ
  \frac{1}{\tau_{6\nu\kappa}}  = C_6\urms k_1+ \Cnukap(\nu+\kappa)\frac{k_1^2}{2}, \qquad \label{eq:scal_compl}
  \frac{1}{\tau_{7\kappa\kappa}}  = C_7\urms k_1  + \Ckapkap\kappa k_1^2.
\EN
Multiplying by the viscous time $\tauvisc= (\nu k_1^2)^{-1}$    yields 
\EQ
  \frac{1}{\tilde{\tau}_{6\nu\kappa}}  = C_6 s \Rey +  \Cnukap,\qquad
  \frac{1}{\tilde{\tau}_{7\kappa\kappa}}  = C_7s  \Pe + \Ckapkap , \label{eq:linscal}
\EN
where a tilde indicates normalization by $\tauvisc$.
Hence, when considering $1/\tilde{\tau}_{6\nu\kappa}$ and
$1/\tilde{\tau}_{7\kappa\kappa}$ as functions of $\Rey$ (or $\Pe$), the wanted parameters should be obtainable by a linear regression analysis.
Figure \ref{fig:fit_Sc1} shows both functions \eq{eq:linscal} for different values of $s$. Obviously, a linear relation is clearly present both for large and small values of $\Rey$,
but with very different fit parameters for the two ranges.
\begin{figure}[t!]\begin{center}
\includegraphics[width=\linewidth]{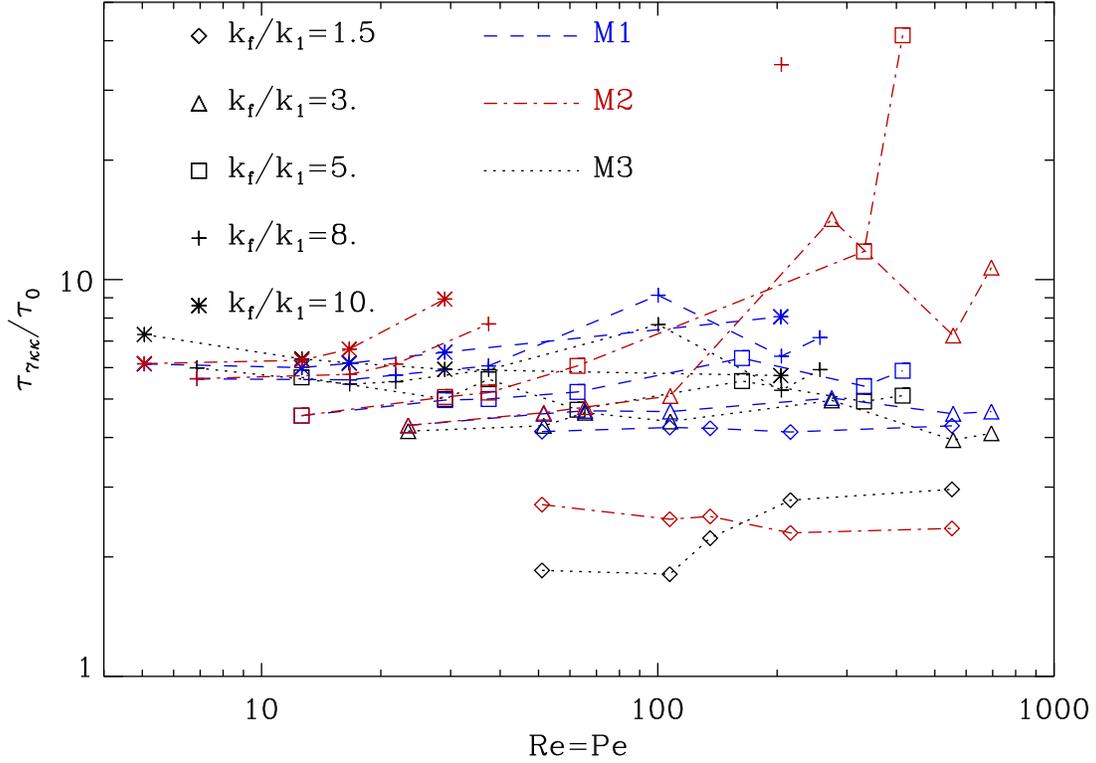}
\end{center}\caption[]{ \label{fig:rect7}
Comparison of relaxation timescales $\tau_{7 \kappa \kappa}$
from methods M1 -- M3.
Negative results are not plotted.
For explanations see \Fig{fig:rect6}.
}\end{figure}
\begin{figure}
\includegraphics[width=\linewidth]{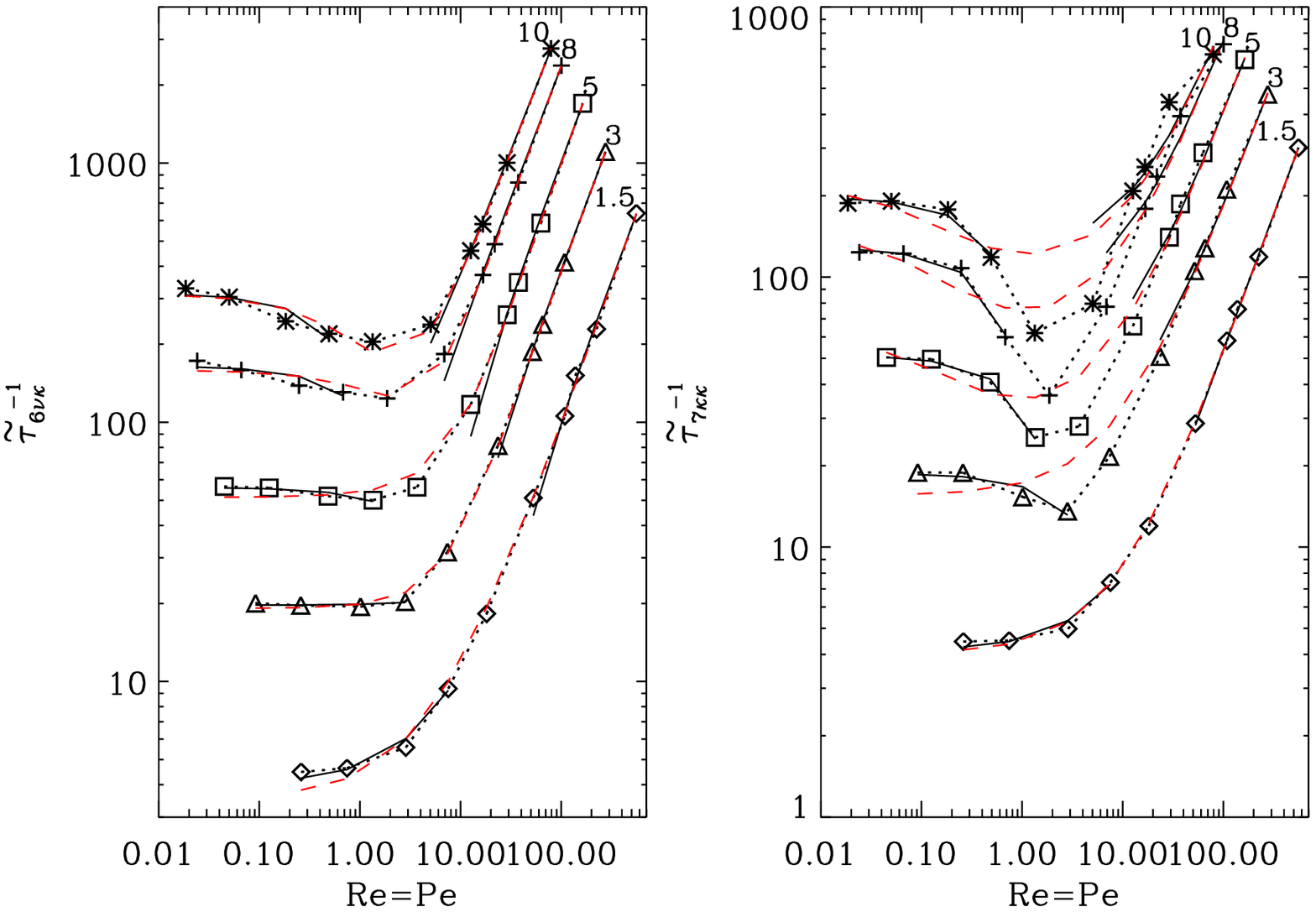}
\caption{\label{fig:fit_Sc1} Relaxation times $\tilde{\tau}_{6\nu\kappa}$ (left) and $\tilde{\tau}_{7\kappa\kappa}$ (right), normalized to the viscous time $\tauvisc$ as functions of $\Rey=\Pe$,
for different values of the scale separation $s$, indicated at the curves.   
Dotted lines with symbols: data from method M1; solid lines:  linear fits according to \eq{eq:linscal}, separately for low and high $\Rey$.
Red dashed: approximation by \eq{eq:newscal} with parameters $C_6$, $C_6'$, $\Cnukap$, $\Cnukap'$ from a best fit,
see \Tab{tab:fitpars}.}
\end{figure}
Guided by these functional dependencies 
we hence propose as an alternative for \eq{eq:linscal} 
\EQ
  \frac{1}{\tilde{\tau}_{6\nu\kappa}} = C_6 s \Rey + \Cnukap+ \frac{1}{C_6' s \Rey + \Cnukap'} \equiv F(\Rey),  \label{eq:newscal}
\EN
and analogously for $1/\tilde{\tau}_{7\kappa\kappa}$.
This ansatz allows to model  linear dependencies on $\Rey$ both for small and large arguments, but with different coefficients:
\begin{eqnarray}
  F(\Rey) &\approx C_6 s \Rey  + \Cnukap \quad &\mbox{for} \quad \Rey \rightarrow \infty , \\
  F(\Rey) &\approx \left(C_6 - \frac{C_6'}{{\Cnukap'}^2} \right) s \Rey + \Cnukap + \frac{1}{\Cnukap'}   \quad &\mbox{for} \quad \Rey \rightarrow 0
\end{eqnarray}
where the slopes may well be different in sign.
The constants in \eq{eq:newscal} were determined by a standard fitting procedure and are given in  \Tab{tab:fitpars}. 
For $\tilde{\tau}_{6\nu\kappa}$ the fit is surprisingly good
throughout, whereas for $\tilde{\tau}_{7\kappa\kappa}$ this holds true
merely for $s=1.5$.
For larger scale separation the ansatz fails to model the pronounced
minima at $\Rey\approx1$ visible in \Fig{fig:fit_Sc1} (right panel).
\begin{table}
\caption{\label{tab:fitpars}Fit parameters of the scaling \eq{eq:newscal} for the results shown in \Fig{fig:fit_Sc1}. }       
\begin{tabular*}{\textwidth}{ccccccccc}
\br
s & $C_6$ & $\Cnukap$ & $10^5 C_6'$ & $10^{3} \Cnukap'$  &  $C_6 - C_6'/{\Cnukap'}^{\!\!\!2}$ &  $\Cnukap + 1/\Cnukap'$ & $C_7$ & $\Ckapkap$ \\
\mr
1.5  & 0.86   &   -121.04  &  1.93 &  8.02 &  0.56  &  3.59  & 0.36        &  -0.20\\
3     & 1.39   &     -45.21  &  29.23 &  15.5    &  0.17  &  19.31 & 0.56       & 15.59 \\
5     & 2.26   &   -177.86  &  3.47 &  4.36 &  0.44  &  51.31  & 0.75      & 36.01 \\
8     &  3.06  &   -88.07    &  12.47 &  4.05 & -4.54    &  158.69   & 0.84 & 76.88 \\
10   & 3.50  &   -13.25    &   25.97 &  3.08 & -23.86 &  311.38   & 0.74 &  121.19\\
\br
\end{tabular*}
\end{table}

The slopes for high $\Rey$, that is, $C_6$ and $C_7$, show a possible
saturation with growing scale separation $s$, but the other coefficients
do not.
Hence, universality of the ansatz \eq{eq:newscal} with respect to
scale separation is questionable.

\subsection{Comparison of methods}

Let us next compare the relaxation timescales $\tau_{6\nu\kappa}$ and
$\tau_{7\kappa\kappa}$ obtained with the different methods 
in more detail.

\subsubsection{Methods M1 and M3.}
\begin{figure}
\hspace*{-8mm}\includegraphics[width=.56\linewidth]{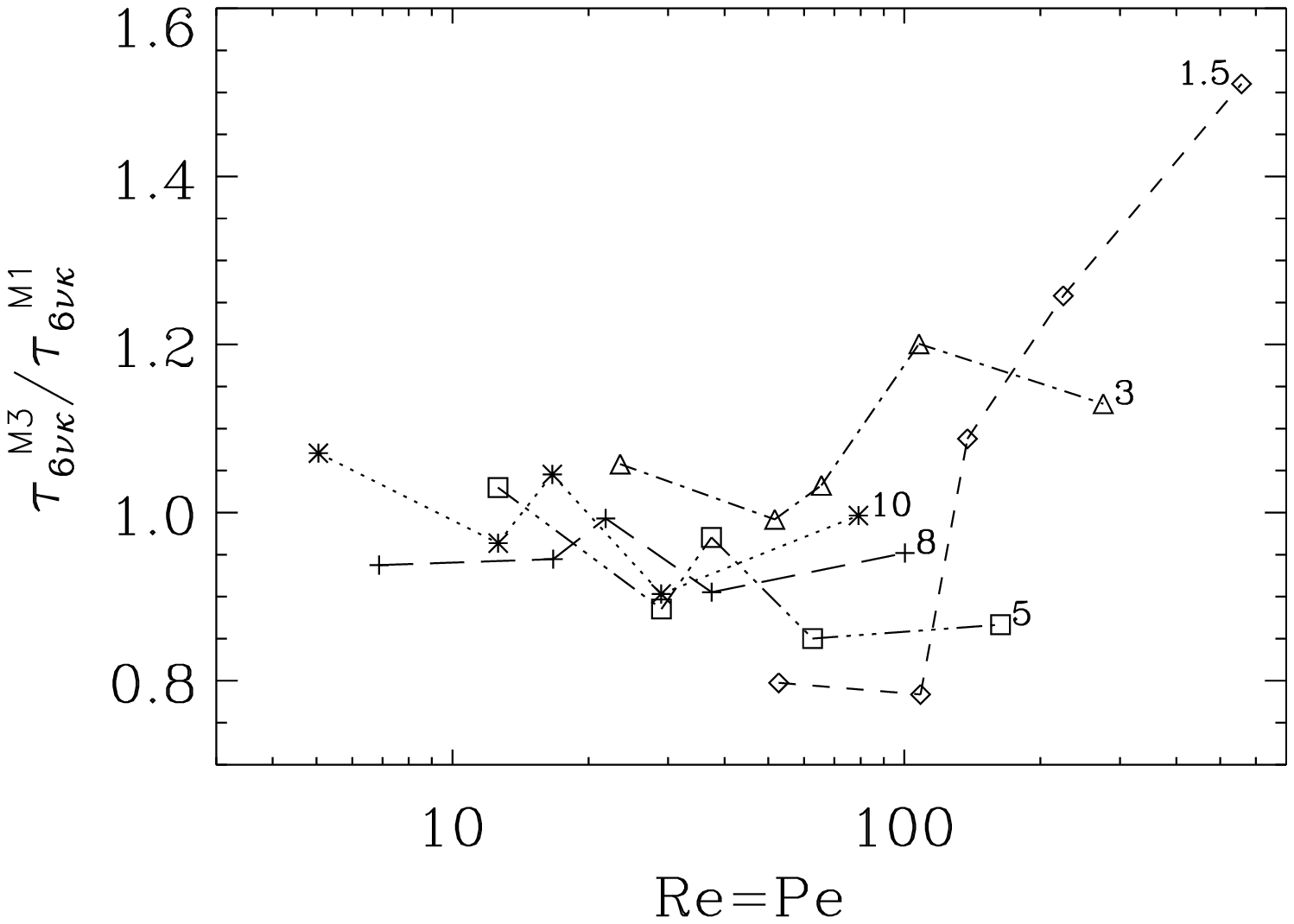}\hspace{-7mm}
\includegraphics[width=.56\linewidth]{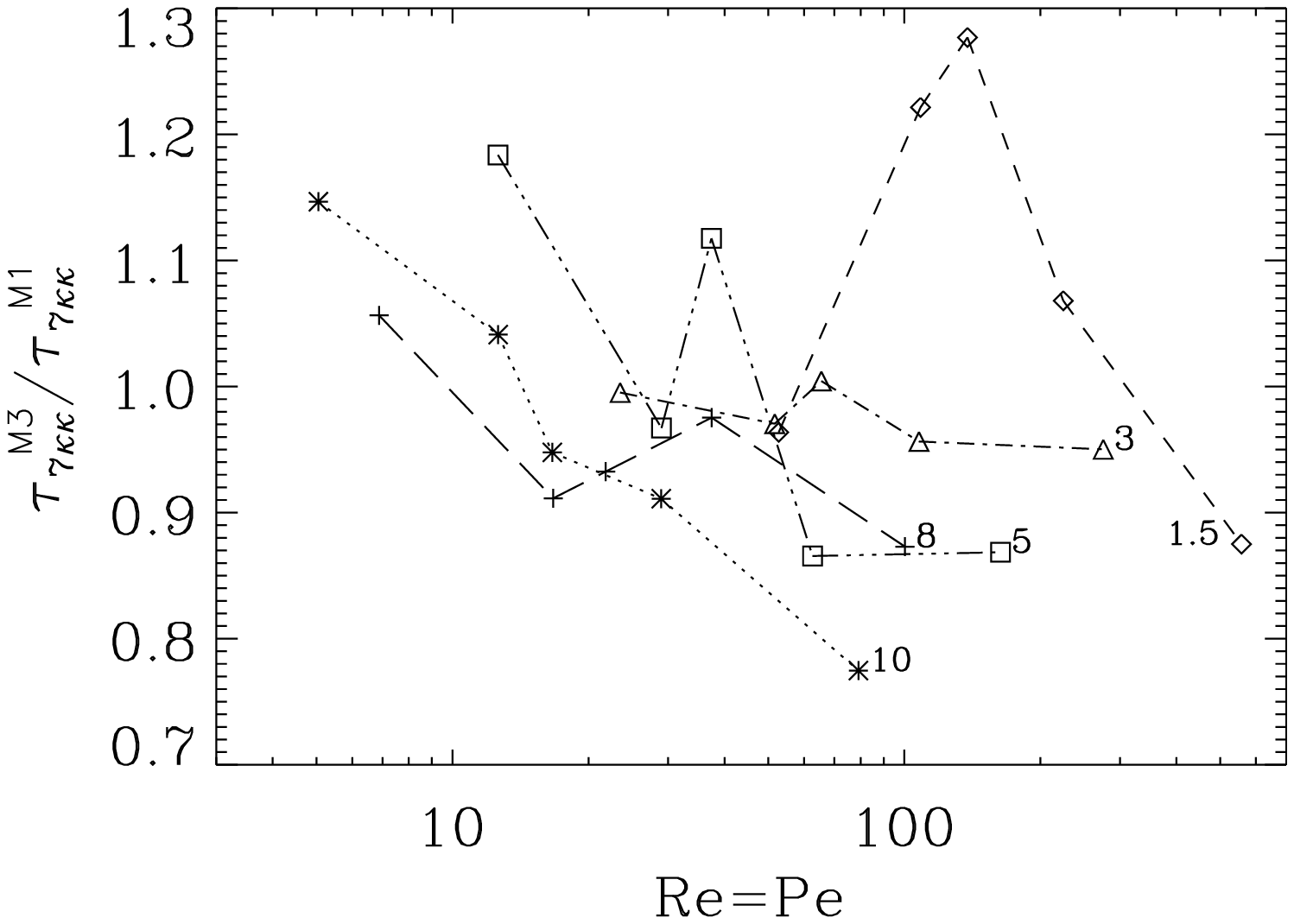}
\caption{\label{fig:comp13} Ratio of the values of $\tau_{6\nu\kappa}$ (left) and 
$\tau_{7\kappa\kappa}$ (right) determined by methods M1 and M3.
Labels indicate the degree of scale separation $s$.}
\end{figure}

\FFig{fig:comp13} shows the ratio of the values of $\tau_{6\nu\kappa}$ and $\tau_{7\kappa\kappa}$ determined
by either of the two methods in dependence on $\Rey$ and $s$.  The M3 values differ from the M1 ones by up to $\pm 30$ \%.
For $\tau_{6\nu\kappa}$ the deviations generally diminish with
decreasing $\Rey$ and with growing scale
separation, falling beneath 10\% for $s=8,10$. However, neither of these tendencies can be confirmed for  $\tau_{7\kappa\kappa}$.

In view of the discussion in \Sec{sect:assess} it is satisfactory to find improving agreement between M1 and M3
with increasing scale separation for which our forced turbulence is more
and more approaching the desired target
of isotropic stationary turbulence.
%
%

\subsubsection{Methods M1 and M2.}

\begin{figure}
\hspace*{-8mm}\includegraphics[width=.56\linewidth]{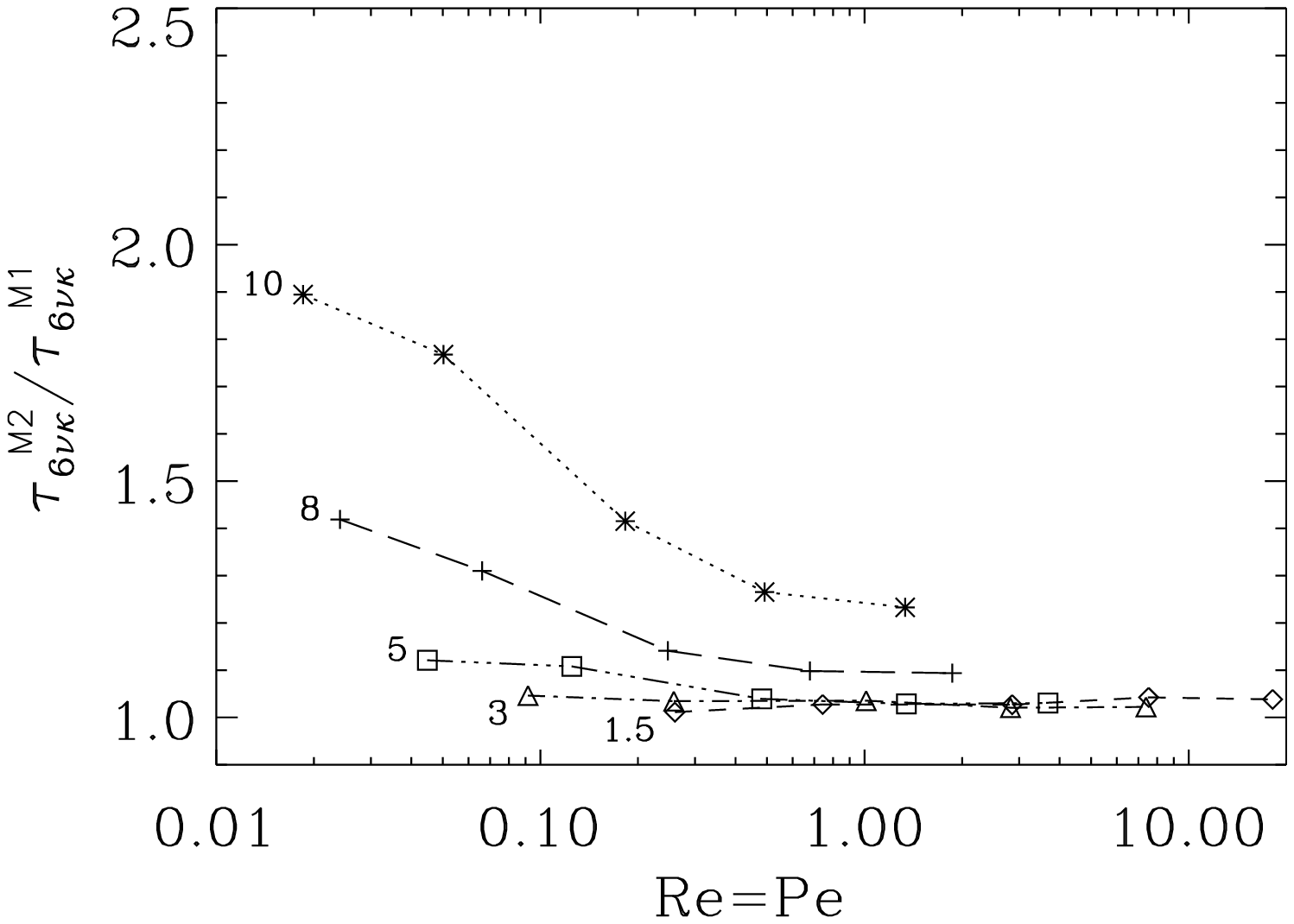}\hspace{-7mm}
\includegraphics[width=.56\linewidth]{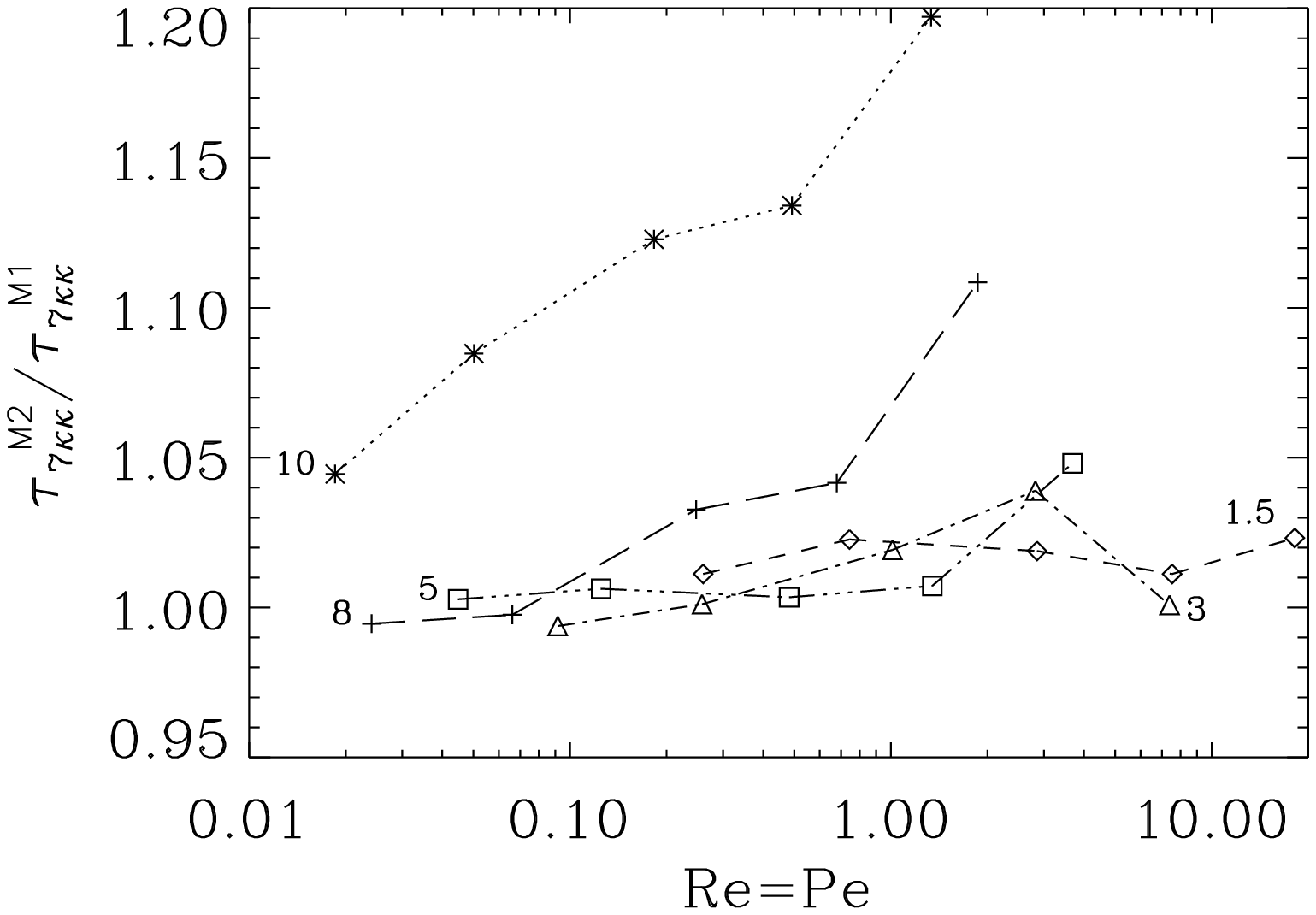}
\caption{\label{fig:comp12} Ratio of the values of $\tau_{6\nu\kappa}$ (left) and 
$\tau_{7\kappa\kappa}$ (right) determined by methods M1 and M2.
Labels indicate the degree of scale separation $s$.}
\end{figure}

As can be seen in \Fig{fig:comp12}, for  small scale separations $s=1.5,3,5$ 
the values of $\tau_{6\nu\kappa}$ from both methods coincide 
fairly. 
Deviations lie within errors.
For large $s$, $s=8$, 10, however, we find the deviations grow with
falling $\Rey$, being still within 
errors
around $\Rey=1$.
We have to conclude that the neglect of $\mean{c\ff}$ has its strongest effects for low $\Rey$ and high $s$.
In contrast, the differences between the  $\tau_{7\kappa\kappa}$ values
from methods M1 and M2 are much smaller, reaching a significant magnitude
(exceeding 
errors
) only for the higher $s=8,10$ and $\Rey\gtrsim1$.
A possible reason for this is insufficient numerical
resolution.
%
%

\section{Comments and extensions}

\subsection{Alternative scaling}
\begin{figure}
\includegraphics[width=\linewidth]{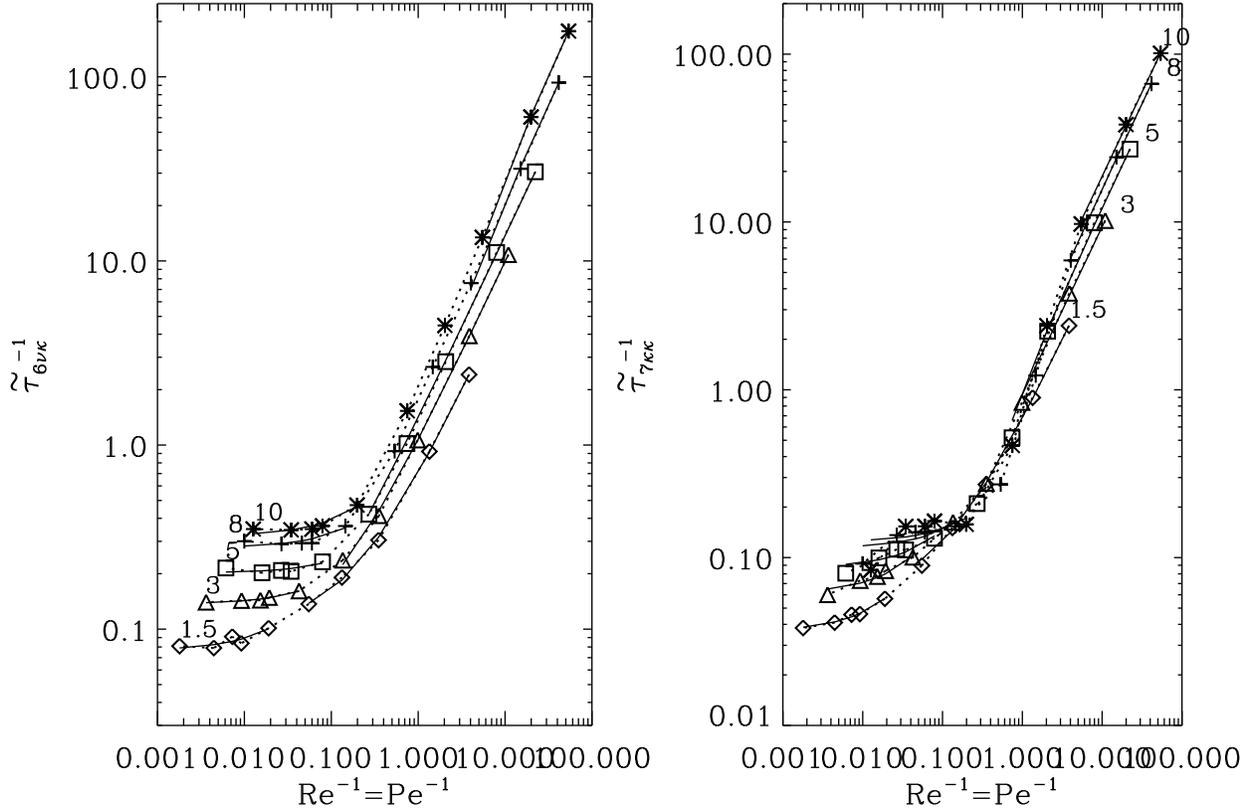}
\caption{\label{fig:fit_Sc1_alt} Relaxation times $\tilde{\tau}_{6\nu\kappa}$ (left) and $\tilde{\tau}_{7\kappa\kappa}$ (right), normalized on the dynamical 
time $\tau_0$  as functions of $\Rey^{-1}=\Pe^{-1}$,
for different values of the scale separation $s$, indicated at the curves.  
Dotted lines with symbols: data from method M1; solid lines:  linear fits according to \eq{eq:scal_alt}, separately for low and high $\Rey$.}
\end{figure}

As an alternative to \eq{eq:scal_compl} one might consider
\EQ
  \frac{1}{\tau_{6\nu\kappa}}  = C_6\urms \kf+ \Cnukap(\nu+\kappa)\frac{\kf^2}{2}, \qquad 
  \frac{1}{\tau_{7\kappa\kappa}}  = C_7\urms \kf  + \Ckapkap\kappa \kf^2.
\EN
Then by multiplying with the dynamic time $\tau_0 = (\urms\kf)^{-1}$ we arrive at
\EQ
  \frac{1}{\tilde{\tau}_{6\nu\kappa}}  = C_6 +  \frac{\Cnukap}{\Rey}, \qquad \label{eq:scal_alt}
  \frac{1}{\tilde{\tau}_{7\kappa\kappa}}  = C_7  + \frac{\Ckapkap}{\Pe}.
\EN
where the normalization is now with respect to $\tau_0$.
Figure \ref{fig:fit_Sc1_alt} shows the same results as \Fig{fig:fit_Sc1}, but now with the altered scaling. Again, a linear fit is viable on each curve, but only separately
for low and high $\Rey$. The corresponding fit parameters can be found in \Tab{tab:fitpars_alt}. Unfortunately, an overall fit analogous to \eq{eq:newscal} does here not work satisfactorily.
\begin{table}
\caption{\label{tab:fitpars_alt}

Alternative fit
parameters of the scaling \eq{eq:scal_alt} for the results shown in \Fig{fig:fit_Sc1_alt}. } 
\begin{center}
\begin{tabular*}{.77\textwidth}{c@{\hspace{8mm}}cccc@{\hspace{9mm}}cccc}
\br
s & \multicolumn{4}{c}{small $\Rey$}Ê& \multicolumn{4}{c}{large $\Rey$}\\      
   & $C_6$ & $\Cnukap$ & $C_7$ & $\Ckapkap$  & $C_6$ & $\Cnukap$ & $C_7$ & $\Ckapkap$ \\
\mr
1.5  & 0.10     &    0.60   &0.064 &     0.61   & 0.077   & 1.25    &0.037 &      1.08\\
3     &  0.091  &    0.98   &0.0025&      0.94 & 0.14      & 0.55   & 0.061 &      0.97\\
5     &  0.052  &    1.37   &-0.26 &       1.23  & 0.20      & 0.32   & 0.088 &      0.59\\
8     &  -2.21   &    2.29   &-0.43 &       1.62  &  0.28     & 0.53   & 0.11 &      0.33\\
10   & -5.84    &   3.39    &-0.11 &       1.88  &  0.32     & 0.72   & 0.12 &      0.23\\
\br
\end{tabular*}
\end{center}
\end{table}

\subsection{ 
An Ogilvie
approach for compressible hydrodynamics?}
One could be tempted to treat the system \eqs{eq:Ccomp}{eq:Ucomp}  in the spirit of the Ogilvie approach quite analogously to what was shown in \Sec{sec:ideal}.
In the ideal case and with $\meanUU=\zervec$, $\meanH=0$ we have for the fluctuating fields
\begin{eqnarray}
\parder{\uu}{t} &=& - \big(\uu \cdot \nab\uu\big)' - \nab h + {\ff}, \label{eq:Ucompfluct}\\
\parder{h}{t} &=& -(\uu\cdot\nab h)' - \cs^2 \nab \cdot \uu,\label{eq:Hcompfluct}
\end{eqnarray}
and for the mean flux an analogue to \eq{eq:Fmean}, but with the term  $-\overline{c\nab p}/\rho$ replaced by 
$-\overline{c\nab h}$.
To close the system, an evolution equation for the quantity $\overline{c\nab h}$ seems hence to be indicated.
From \eq{eq:Cfluct} and \eq{eq:Hcompfluct} we get
\begin{eqnarray}
\hspace*{-1cm}\parder{\,\overline{c\nab h}}{t} &=& -\cs^2\overline{c\nab \nab \cdot \uu } - \overline{c\nab(\uu\cdot\nab h)} 
                                                                                           -\overline{\nab h \nab\cdot(\uu \meanC)} -\overline{\nab h\cdot \nab(\uu c)}, \\
\hspace*{-1cm}\parder{\left(\overline{c\nab h}\right)_i}{t} &=&  -\overline{ \parder{h}{x_i}\parder{u_j}{x_j}}\meanC -\overline{u_j\parder{h}{x_i}}\parder{\meanC}{x_j} 
- \cs^2 \overline{c\parder{\,^2u_j}{x_i \partial x_j} } - \;\mbox{third order terms},
\end{eqnarray}
where the second order correlation $\overline{c\,\parderil{^2u_j}{x_i \partial x_j}}$ can only partly be expressed by $\meanFFF_i$. 
The remaining parts could be modelled by a $\tau$ ansatz as used for the diffusion terms in \Sec{sec:nonideal}, but note that here the ``diffusivity''
is
$\cs^2$
 and we have no argument to consider the not properly modelled terms as small.

\section{Conclusions}

The main conclusion to be drawn from the present work is that the
time scales used to model closure terms in the equations for the
mean
 flux
$\overline{\uu c}$ and the mean square concentration $\overline{c^2}$
are nearly independent of $\Rey$ for $\Rey\geq10$ and also nearly
independent of the scale separation ratio for $\kf/k_1\ge3$.
Expressed in terms of
dynamical
 times, the resulting non-dimensional time
scales 
can be
referred to as Strouhal numbers whose values are around 3 for
the $\overline{\uu c}$ closure term and around 7 for the $\overline{c^2}$
closure term.
The former
value
 is in good agreement with earlier work using the
$\tau$ approximation \cite{BKM04}.

Equipped with this knowledge, we may now be better justified in using
the closure hypotheses discussed here for the quantities
$\overline{\uu c}$ and $\overline{c^2}$.
On the other hand, as explained in the present paper, it is quite clear
that these closure hypotheses lack thorough justification \cite{RR07}.
One should therefore in future strive to find systematic discrepancies
from the anticipated scalings.
One example that we alluded to in the present paper is the inhomogeneous
case in which the $\tau$ approach my break down.
Future work in that direction seems now to be highly desirable,
because in virtually all astrophysical applications the turbulence is
inhomogeneous or at least anisotropic.

\section*{Acknowledgements}

The authors acknowledge the hospitality of NORDITA during the
program `Dynamo, Dynamical systems and Topology.' 
We acknowledge the allocation of computing resources by CSC - IT
Center for Science Ltd.\ in Espoo, Finland, who are administered by
the Finnish Ministry of Education. This work was supported in part by
the Finnish Cultural Foundation (JES), the Finnish Academy grants
121431, 136189 (PJK, MR) and 218159, 141017 (MJK), the European
Research Council under the AstroDyn Research Project 227952, and
the Swedish Research Council under Project 621-2011-5076.

\vspace{2cm}
\bibliographystyle{plain}
\bibliography{paper}

\begin{thebibliography}{10}

\bibitem{BF02}
E.~G. {Blackman} and G.~B. {Field}.
\newblock {New Dynamical Mean-Field Dynamo Theory and Closure Approach}.
\newblock {\em Physical Review Letters}, 89(26):265007, 2002.

\bibitem{BF03}
E.~G. {Blackman} and G.~B. {Field}.
\newblock {A new approach to turbulent transport of a mean scalar}.
\newblock {\em Physics of Fluids}, 15:L73--L76, 2003.

\bibitem{B01}
A.~{Brandenburg}.
\newblock {The Inverse Cascade and Nonlinear Alpha-Effect in Simulations of
  Isotropic Helical Hydromagnetic Turbulence}.
\newblock {\em \apj}, 550:824--840, 2001.

\bibitem{BHB11}
A.~{Brandenburg}, N.~E.~L. {Haugen}, and N.~{Babkovskaia}.
\newblock {Turbulent front speed in the Fisher equation: Dependence on
  Damk{\"o}hler number}.
\newblock {\em \pre}, 83(1):016304, 2011.

\bibitem{BKM04}
A.~{Brandenburg}, P.~J. {K{\"a}pyl{\"a}}, and A.~{Mohammed}.
\newblock {Non-Fickian diffusion and tau approximation from numerical
  turbulence}.
\newblock {\em Physics of Fluids}, 16:1020--1027, 2004.

\bibitem{BN11}
A.~{Brandenburg} and {\AA}.~{Nordlund}.
\newblock {Astrophysical turbulence modeling}.
\newblock {\em Rep. Progr. Phys.}, 74(4):046901, 2011.

\bibitem{BS05}
A.~{Brandenburg} and K.~{Subramanian}.
\newblock {Minimal tau approximation and simulations of the alpha effect}.
\newblock {\em \aap}, 439:835--843, 2005.

\bibitem{BS07}
A.~{Brandenburg} and K.~{Subramanian}.
\newblock {Simulations of the anisotropic kinetic and magnetic alpha effects}.
\newblock {\em Astronomische Nachrichten}, 328:507, 2007.

\bibitem{GO05}
P.~{Garaud} and G.~I. {Ogilvie}.
\newblock {A model for the nonlinear dynamics of turbulent shear flows}.
\newblock {\em J. Fluid Mech.}, 530:145--176, 2005.

\bibitem{GOMS10}
P.~{Garaud}, G.~I. {Ogilvie}, N.~{Miller}, and S.~{Stellmach}.
\newblock {A model of the entropy flux and Reynolds stress in turbulent
  convection}.
\newblock {\em \mnras}, 407:2451--2467, 2010.

\bibitem{HB09}
A.~{Hubbard} and A.~{Brandenburg}.
\newblock {Memory Effects in Turbulent Transport}.
\newblock {\em \apj}, 706:712--726, 2009.

\bibitem{KB08}
P.~J. {K{\"a}pyl{\"a}} and A.~{Brandenburg}.
\newblock {Lambda effect from forced turbulence simulations}.
\newblock {\em \aap}, 488:9--23, 2008.

\bibitem{KR80}
F.~{Krause} and {K.-H.} {R\"adler}.
\newblock {\em {Mean-field magnetohydrodynamics and dynamo theory}}.
\newblock Pergamon Press, Ltd., Oxford, 1980.

\bibitem{LKKBL09}
A.~J. {Liljestr{\"o}m}, M.~J. {Korpi}, P.~J. {K{\"a}pyl{\"a}},
  A.~{Brandenburg}, and W.~{Lyra}.
\newblock {Turbulent stresses as a function of shear rate in a local disk
  model}.
\newblock {\em Astron. Nachr.}, 330:92--99, 2009.

\bibitem{MB10}
E.~J.~M. {Madarassy} and A.~{Brandenburg}.
\newblock {Calibrating passive scalar transport in shear-flow turbulence}.
\newblock {\em \pre}, 82(1):016304, 2010.

\bibitem{MG07}
N.~{Miller} and P.~{Garaud}.
\newblock {A Practical Model of Convective Dynamics for Stellar Evolution
  Calculations}.
\newblock In {R.~J.~Stancliffe, G.~Houdek, R.~G.~Martin, \& C.~A.~Tout},
  editor, {\em Unsolved Problems in Stellar Physics: A Conference in Honor of
  Douglas Gough}, volume 948 of {\em Am. Inst. Phys. Conf. Ser.}, pages
  165--169, 2007.

\bibitem{M78}
H.~K. {Moffatt}.
\newblock {\em {Magnetic field generation in electrically conducting fluids}}.
\newblock Cambridge University Press, Cambridge, England, 1978.

\bibitem{O03}
G.~I. {Ogilvie}.
\newblock {On the dynamics of magnetorotational turbulent stresses}.
\newblock {\em \mnras}, 340:969--982, 2003.

\bibitem{RR07}
K.-H. {R{\"a}dler} and M.~{Rheinhardt}.
\newblock {Mean-field electrodynamics: critical analysis of various analytical
  approaches to the mean electromotive force}.
\newblock {\em Geophys. Astrophys. Fluid Dyn.}, 101:117--154, 2007.

\bibitem{RB11}
M.~{Rheinhardt} and A.~{Brandenburg}.
\newblock {Modeling spatio-temporal nonlocality in mean-field dynamos}.
\newblock {\em Astron. Nachr.}, 333:80--86, 2012.

\bibitem{R89}
G.~{R\"udiger}.
\newblock {\em {Differential rotation and stellar convection. Sun and the solar
  stars}}.
\newblock Akademie-Verlag, Berlin, 1989.

\bibitem{RH04}
G.~{R{\"u}diger} and R.~{Hollerbach}.
\newblock {\em {The magnetic universe: Geophysical and Astrophysical Dynamo
  Theory}}.
\newblock Wiley-VCH, Berlin, 2004.

\bibitem{SBKM11}
J.~E. {Snellman}, A.~{Brandenburg}, P.~J. {K{\"a}pyl{\"a}}, and M.~J.
  {Mantere}.
\newblock {Verification of Reynolds stress parameterizations from simulations}.
\newblock {\em Astron. Nachr.}, 332:883--888, 2011.

\bibitem{SKKL09}
J.~E. {Snellman}, P.~J. {K{\"a}pyl{\"a}}, M.~J. {Korpi}, and A.~J.
  {Liljestr{\"o}m}.
\newblock {Reynolds stresses from hydrodynamic turbulence with shear and
  rotation}.
\newblock {\em \aap}, 505:955--968, 2009.

\end{thebibliography}

\end{document}